\documentstyle[epsfig,preprint,prl,aps]{revtex}
\tightenlines
\begin{document}
\draft
\title{Test of quantum nonlocality for cavity fields} 

\author{M. S. Kim$^{1,2}$ and Jinhyoung
  Lee$^1$\footnote{hyoung@quanta.sogang.ac.kr} }

\address{$^1$ Department of Physics, Sogang University, CPO Box 1142,
  Seoul 100-611, Korea \\
  $^2$ Department of Applied Mathematics and Theoretical Physics, The
  Queen's University of Belfast, Belfast BT7 1NN, United Kingdom}

\date{\today} 

\maketitle

\begin{abstract}  
  There have been studies on formation of quantum-nonlocal states in
  spatially separate two cavities.  We suggest a nonlocal test for the
  field prepared in the two cavities.  We couple classical driving
  fields with the cavities where a nonlocal state is prepared.  Two
  independent two-level atoms are then sent through respective
  cavities to interact off-resonantly with the cavity fields. The
  atomic states are measured after the interaction.  Bell's inequality
  can be tested by the joint probabilities of two-level atoms being in
  their excited or ground states.  We find that quantum nonlocality
  can also be tested using a single atom sequentially interacting with
  the two cavities.  Potential experimental errors are also
  considered.  We show that with the present experimental condition of
  5\% error in the atomic velocity distribution, the violation of
  Bell's inequality can be measured.
\end{abstract}
\vspace{1.5truecm}

\newpage
\section{Introduction}

Entangled states have been at the focus of discussions in quantum
information theory encompassing quantum teleportation, computing, and
cryptography.  Two-body entanglement \cite{qucor} allows diverse
measurement schemes which can admit tests of quantum nonlocality
\cite{Bell,chsh69}. Using the atom-field interaction in a high-$Q$ cavity
we can produce quantum entanglement between cavity fields, between
atoms, and between a cavity and an atom.
An entangled pair of atoms have been experimentally generated using the
cavity QED (Quantum electrodynamics) \cite{Hagley97}.  
The entanglement of atoms and fields
in the cavity can be utilized towards a realization of the controlled-NOT
gate for quantum computation \cite{Brune94,Davidovich94}.

A pair of atoms can be prepared in an entangled state  
using the atom-field interaction in a high-$Q$ cavity.
The interaction of a single two-level atom with a 
cavity field brings about entanglement of the atom and the cavity field 
\cite{Phoenix93}.  If
the atom does not decay into other internal states after it comes out from
the cavity, the entanglement will
survive for long and it can be transferred to a second atom
interacting with the cavity field.  The violation of Bell's inequality
can be tested by the joint measurement of atomic states.

There are proposals to entangle fields in two spatially
separated cavities using the atom-field interaction
\cite{Meystre92,Davidovich93}.  A two-level atom in its excited state
passes sequentially through two resonant single-mode vacuum cavities
and is found to be in its ground state after the second-cavity
interaction.  If the interaction with the first cavity is equivalent
to a $\pi/2$ vacuum pulse and the second-cavity interaction is to a $\pi$
pulse \cite{Davidovich94}
then the atom could have deposited a photon either in the first
cavity or in the second so that the final state $|\Psi_f\rangle$ of
the two cavity field is \cite{Meystre92}
\begin{equation}
\label{Meystre}
|\Psi_f\rangle = {1 \over \sqrt{2}}(|1,0\rangle + 
\mbox{e}^{i\varphi}|0,1\rangle)
\end{equation}
where $|1,0\rangle$ denotes one photon in the first cavity and none in
the second, and $|0,1\rangle$ vice versa.  Using the entangled cavity
field (\ref{Meystre}), an unknown atomic quantum state can also be
teleported \cite{Davidovich94}.

A three-level atom can act as a quantum switch to couple a vacuum
cavity with the external classical coherent field.  If the atom is in
its intermediate state before it enters to the cavity, the AC Stark
shift between the intermediate and upper-most states brings about the
resonant coupling of the cavity with the external field so that the
external field can be fed into the cavity.  If the atom is initially
in its lower-most state, the atom is unable to switch on the coupling
between the cavity and the external field.  Using the Ramsey
interference and atomic quantum switch, Davidovich {\it et al.}
suggested a coherent state entanglement $|\Psi_c\rangle$ between two
separate cavities \cite{Davidovich93}:
$|\Psi_c\rangle = B_1 |\alpha, 0\rangle + B_2 |0, \alpha\rangle$,
where $|\alpha, 0\rangle$ denotes the first cavity in the coherent
state of the amplitude $\alpha$ and the second in the vacuum.

In this paper, we are interested in a test of nonlocality for the
entangled field prepared in the spatially separated cavities.  Despite
the suggestions on the production of entangled cavity fields, the test
of quantum entanglement, {\it i.e.}, the measurement of the
violation of Bell's inequality for the entangled cavity fields has not
been studied.  To test quantum nonlocality, we first couple classical driving
fields with the cavities where a nonlocal state is prepared.
Two independent two-level atoms are then sent through respective cavities
to interact off-resonantly with the cavity fields. The atomic states
are measured after the interaction.
Bell's inequality can be tested by the joint probabilities of
two-level atoms being in their excited 
or ground states.  We find that quantum nonlocality can also be
tested using a single atom sequentially interacting with the two
cavities. We also consider potential experimental 
errors.  The atoms normally have the Gaussian velocity
distribution with the normalized standard deviation less than 5\%.
We show that, even with the experimental errors caused by the velocity
distribution, the test can be feasible.

\section{Bell's inequality by parity measurement}
\label{sec2}

It is important to choose the type of measurement variables when
testing nonlocality for a given state.  Banaszek and W\'{o}dkiewicz
\cite{Banaszek} considered even and odd parities of the field state 
as the measurement
variables, where a state is defined to be in the even (odd) parity if
the state has even (odd) numbers of photons.  
The even and odd parity operators, $\hat{O}_E$ and $\hat{O}_O$, are
the projection operators to measure the probabilities of the field
having even and odd numbers of photons, respectively:
\begin{equation}
\label{even-operator}
\hat{O}_E = \sum_{n=0}^\infty |2n\rangle \langle 2n|~~~;~~~
\hat{O}_O = \sum_{n=0}^\infty |2n+1\rangle \langle 2n+1|.
\end{equation}

To test the quantum nonlocality for the field state of the modes $a$ and
$b$, we define the quantum correlation operator based on the joint
parity measurements:
\begin{equation}
  \label{eq:tpcf}
  \hat{\Pi}^{ab}(\alpha, \beta) =
  \hat{\Pi}_E^{a}(\alpha)\hat{\Pi}_E^{b}(\beta)
 -\hat{\Pi}_E^{a}(\alpha)\hat{\Pi}_O^{b}(\beta)
 -\hat{\Pi}_O^{a}(\alpha)\hat{\Pi}_E^{b}(\beta)
 +\hat{\Pi}_O^{a}(\alpha)\hat{\Pi}_O^{b}(\beta)
\end{equation}
where the superscripts $a$ and $b$ denote the field modes and the
displaced parity operator, $\hat{\Pi}_{E,O}(\alpha)$, is defined as
\begin{equation}
  \label{eq:dpo}
  \hat{\Pi}_{E,O}(\alpha) = \hat{D}(\alpha) \hat{O}_{E,O} 
  \hat{D}^\dagger(\alpha).
\end{equation}
The displacement operator $\hat{D}(\alpha)$ displaces a
state by $\alpha$ in phase space.  The displaced parity operator acts
like a rotated spin projection operator in the spin measurement
\cite{Bell}.  We can easily derive that the local hidden variable
theory imposes the following Bell's inequality \cite{Banaszek}
\begin{equation}
\label{inequality1}
|B(\alpha,\beta)|\equiv
|\langle \hat{\Pi}^{ab}(0, 0)+\hat{\Pi}^{ab}(0, \beta)
+\hat{\Pi}^{ab}(\alpha, 0)-\hat{\Pi}^{ab}(\alpha, \beta)
\rangle|\leq 2
\end{equation}
where we call $B(\alpha,\beta)$ as the Bell function.

\section{Parity measurement in cavity QED}
Englert {\it et al.} \cite{Englert93} 
proposed an experiment to determine the parity of the
field in a high-Q single-mode cavity.  Let us consider a far-off resonant
interaction of a two-level atom with a single-mode cavity field.  If
the detuning $\Delta=\omega_o-\omega$ of the atomic transition frequency
$\omega_o$ from the cavity-field frequency $\omega$ is much larger
than the Rabi frequency $\Omega$, there is no energy exchange between
the atom and the field but the relative phase of the atomic states
changes due to the AC stark shift \cite{Brune92}.  The change of the
phase depends on the number of photons in the cavity and on the state of 
the atom:
\begin{eqnarray}
\label{phase-change}
|e,\psi_f\rangle &\rightarrow & \exp[-i\Theta(\hat{n}+1)]
|e,\psi_f\rangle,\nonumber \\
|g,\psi_f\rangle &\rightarrow & \exp[i\Theta(\hat{n})]
|g,\psi_f\rangle
\end{eqnarray}
where the atom-field state $|e,\psi_f\rangle$ denotes the atom in its
excited state and the cavity field in $|\psi_f\rangle$.  The phase
$\Theta(\hat n)$ is a function of the number of photons in the cavity
and the atom-field coupling time and strength.

If a $\pi/2$ pulse is
applied on a two-level atom before it enters the cavity, 
the atom initially
in its excited state transits to a superposition state without
changing the cavity field
\begin{equation}
\label{piovertwo}
|e,\psi_f \rangle \rightarrow 
{1\over\sqrt{2}}[|e,\psi_f\rangle +i \mbox{e}^{i\phi_1}
|g, \psi_f \rangle],
\end{equation}
where $\phi_1$ is determined by the phase of the pulse field.  
The atom, then, passes  through a cavity and undergoes an
off-resonant interaction with the cavity field
where the atom-field coupling function is selected to be
\begin{equation}
\label{theta-value}
\Theta(\hat n)={\pi \over 2} \hat n.
\end{equation}
After
the atom comes out from the cavity, the atom is let interact with
the second $\pi/2$ pulse.   
If the phases of the first and the second pulses are chosen to
satisfy the following relation
\begin{equation}
\label{phases-value}
i\mbox{e}^{i(\phi_1 - \phi_2)}=1,
\end{equation}
the atom-field state after the second $\pi/2$ pulse is
\begin{equation}
\label{final-singleatom}
{1 \over 2} \Big\{i[1-(-1)^{\hat n}]|e,\psi_f\rangle 
+\mbox{e}^{i\phi_2}[1+(-1)^{\hat n}]|g,\psi_f\rangle\Big\}.
\end{equation} 
If the atom is detected in its excited state, the field has the even
parity.  If the atom is detected in its ground state, the field has
the odd parity.  The atom tells us the parity of the field
\cite{Englert93}.

\section{Quantum nonlocality of cavity fields}
\label{sec:qncf}

To test quantum nonlocality of the field, $|\psi^{ab}_f\rangle$, prepared
in spatially separated two single-mode cavities, we use two two-level
atoms as shown in Fig.~1(a).  In this paper we assume that the
mode structures of the cavities are identical and 
the atoms are independent identical atoms.  
The atoms are labelled as $a$ and $b$ to
interact, respectively, with the fields in the cavities $a$ and $b$.
Each atom is initially prepared in its excited state and sequentially
passes through interaction zones of the first $\pi /2$ pulse, the cavity
field, and the second $\pi/2$ pulse.  The atoms-field state then
becomes
\begin{equation}
\label{twoatomfield0}
|e_a,e_b\rangle|\psi_f^{ab}\rangle\rightarrow 
|\varphi(\hat{n}_a,\hat{n}_b)\rangle|\psi_f^{ab}\rangle
\end{equation}
where $|\varphi(\hat{n}_a,\hat{n}_b)\rangle$ is the atomic state
with the weights of the field operators ($\hat{n}_a$ and $\hat{n}_b$
are number operators of fields in the cavities $a$ and $b$):
\begin{eqnarray}
  \label{eq:ertwoatom}
  |\varphi(\hat{n_a}, \hat{n_b})\rangle = & &
    a(\hat{n}_a) a(\hat{n}_b)|e_a,e_b\rangle 
  + a(\hat{n}_a) b(\hat{n}_b)|e_a,g_b\rangle \nonumber \\
  &+& b(\hat{n}_a) a(\hat{n}_b)|g_a,e_b\rangle 
  + b(\hat{n}_a) b(\hat{n}_b)|g_a,g_b\rangle 
\end{eqnarray}
where $a(\hat{n})= [e^{-i\Theta(\hat{n}+1)} - e^{i(\phi_1
  -\phi_2)}e^{i\Theta(\hat{n})}]/2$ and
$b(\hat{n})=ie^{i\phi_2}[e^{-i\Theta(\hat{n}+1)} + e^{i(\phi_1
  -\phi_2)}e^{i\Theta(\hat{n})}]/2$.
Choosing appropriate conditions for the atom-field couplings
and pulse phases as shown in (\ref{theta-value})
and (\ref{phases-value}), the atoms-field state becomes
\begin{eqnarray}
\label{twoatomfield}
{1 \over 4}&& \Big\{i[1-(-1)^{\hat{n}_a}]|e_a\rangle +
\mbox{e}^{i\phi_2}[1+(-1)^{\hat{n}_a}]|g_a\rangle\Big\}
\nonumber \\
&&
\times\Big\{i[1-(-1)^{\hat{n}_b}]|e_b\rangle +
\mbox{e}^{i\phi_2}[1+(-1)^{\hat{n}_b}]
|g_b\rangle\Big\}|\psi^{ab}_f\rangle.
\end{eqnarray}
If the atoms are jointly found in their excited states then we know
that  both the cavities are in the odd parity states.  
The joint probability $P_{ee}$ of the atoms being in their excited
states is related to the expectation value of 
the following joint parity operator:
\begin{equation}
\label{jointprob-meanparity}
P_{ee}=\langle \hat{\Pi}_O^a(0)\hat{\Pi}_O^b(0)\rangle.
\end{equation}

In Eqs.~(\ref{eq:tpcf}) and (\ref{inequality1}), it is seen that we
need to know the joint parities of
the {\it displaced} original fields to test 
quantum nonlocality.  To displace the
cavity field, external stable fields are coupled to the cavities as
shown in Fig.~1(a) \cite{Brune96}.  After a nonlocal field state is prepared
in the cavities, we couple the cavities with the external fields to
displace the original nonlocal field, then send two independent atoms through
respective cavities.  The $\pi/2$ pulses shine atoms before and after
the cavity interaction to provide Ramsey interference effects.  The
atomic states are detected after the second $\pi/2$ pulses.
$P_{ee}(\alpha,\beta)$ denotes the joint probability of atoms being in
their excited states when the original fields 
in the cavities $a$ and $b$ are
displaced by $\alpha$ and $\beta$, respectively.  The expectation value of
the quantum correlation operator in (\ref{eq:tpcf}) is obtained by the
joint probabilities:
\begin{equation}
\label{joint-quantumcorrelation}
\langle \hat{\Pi}^{ab}(\alpha,\beta)\rangle 
=P_{ee}(\alpha,\beta) - P_{eg}(\alpha,\beta)- 
P_{ge}(\alpha,\beta) + P_{gg}(\alpha,\beta).
\end{equation}
If there are any displacement factors $\alpha$ and $\beta$ which
result in the violation of Bell's inequality in Eq.~(\ref{inequality1}),
the field originally prepared in the cavities is quantum-mechanically 
nonlocal.

After a closer look at Eq.~(\ref{eq:tpcf}), we find that we do not need 
the individual parity of each cavity field to test the inequality
(\ref{inequality1}).  We need  the parity
of only the total field. 
Instead of sending two atoms to cavities, we now send a single
two-level atom sequentially through cavities as shown 
in Fig.~1(b).  The atom is
initially prepared in its excited state and undergoes  $\pi/2$-pulse
interactions before and after the cavity interaction.  The atom-field
coupling strength is selected to satisfy Eq.~(\ref{theta-value}) and
the phases, $\phi_a$ and $\phi_b$, of the two $\pi/2$ pulses are chosen as
$\exp[i(\phi_a-\phi_b)]=1$ then the atom-field state becomes
\begin{equation}
\label{atomicstate-oneatom}
|e,\psi_f^{ab}\rangle\rightarrow 
-{1\over2}[1+(-1)^{\hat{n}_a+\hat{n}_b}]|e,\psi_f^{ab}\rangle
+{i\over 2} [1-(-1)^{\hat{n}_a+\hat{n}_b}]|g,\psi_f^{ab}\rangle.
\end{equation}
The external stable fields are taken to be 
coupled with the cavities to displace
the cavity fields.  The probability $P_e(\alpha,\beta)$ of the atom
being in its excited state after having passed displaced cavity fields
and $\pi/2$ pulses, is the expectation value of the parity operators:
\begin{equation}
\label{jointprob-meanparity2}
P_{e}(\alpha,\beta)=\langle \hat{\Pi}_O^a(\alpha)\hat{\Pi}_O^b(\beta)
+\hat{\Pi}_E^a(\alpha)\hat{\Pi}_E^b(\beta)\rangle
\end{equation}
where $\alpha$, $\beta$ denote the displacements of the 
fields in the cavities $a$ and $b$. Similarly, the probability of the 
atom being in its ground state $P_g(\alpha,\beta)$
is found to be related to the odd parity of the total fields. 
The expectation value of the quantum correlation function operator in
Eq.~(\ref{eq:tpcf}) is simply
\begin{equation}
\label{joint-quantumcorrelation2}
\langle\hat{\Pi}^{ab}(\alpha,\beta)\rangle = P_e(\alpha,\beta)
-P_g(\alpha,\beta).
\end{equation}
This does not tell us the parity of each mode but the parity of the
total field which is enough to test the violation of Bell's inequality.

\section{Remarks}
We have suggested a simple way to test quantum nonlocality of cavity
fields by measuring the states of atoms after their interaction with
cavity fields.  The test does not require a numerical process on the
measured data.  
The difference in the probability of a single two-level atom 
being in its excited and ground
states is directly related to the test of quantum nonlocality.  
In fact, this can also be used to
reconstruct the two-mode Wigner function as the mean parity of the
field is proportional to the two-mode Wigner function
\cite{Banaszek,Moya93,Lutterbach97}:
\begin{equation}
\label{Wigner-meanparity}
W(\alpha,\beta)=(2/\pi)^2 \langle\hat{\Pi}^{ab}(\alpha,\beta)\rangle.
\end{equation}   

Experimental error can easily occur from the fluctuation in 
the atom-field coupling
strength and time.  
The atom-field coupling function $\Theta(\hat{n})$ depends
on the mode structure of the cavity field and on 
the duration of time for the atom to interact with the cavity
field \cite{Englert93}.  
Because the atomic velocity has some fluctuations the
interaction time is subject to the experimental error
\cite{Kim98}. Another error source may be the $\pi/2$ pulse operation.
We analyze the possibility to measure the violation of quantum
nonlocality in the potential experimental situation. 

The test of quantum nonlocality using the two-atom scheme in Fig. 1(a)
is considered with potential experimental errors.
The error in the atom-field coupling function is denoted
by $\Delta\Theta(\hat{n})$, which is the departure of the experimental
value $\Theta(\hat{n})$ from the required value $\Theta_0(\hat{n})$:
\begin{equation}
  \label{eq:fphaset}
  \Delta\Theta(\hat{n}) = \Theta(\hat{n}) - \Theta_0(\hat{n})=
  \delta\hat{n}
\end{equation}
where $\Theta_0(\hat{n})=(\pi/2)\hat{n}$.  The relative phases of atomic 
states given by the $\pi/2$-pulse interactions are also subject
to experimental errors.  We take the phase error $\Delta\phi$ as
\begin{equation}
\label{phase-error}
\Delta\phi=(\phi_2-\phi_1)-\phi_0~~~;~~~i\exp(i\phi_0)=1.
\end{equation} 
Note that the atomic state measurement 
is equivalent to the parity measurement as in 
Eq.~(\ref{jointprob-meanparity}) only when
$\Theta(\hat{n}) = \Theta_0(\hat{n})$ and $\phi_2-\phi_1 = \phi_0$.

The errors in the atom-field coupling and phases of the $\pi/2$
pulses bring about the departure $\Delta\Pi^{ab}(\alpha,\beta)$ of 
the joint atomic state 
probabilities from the expectation value of the parity operators 
in Eq.~(\ref{joint-quantumcorrelation}).  The mean error of
$\Delta\Pi^{ab}(\alpha,\beta)$ is calculated up to the second order
of $\delta$ and $\Delta\phi$ as follows 
\begin{eqnarray}
\label{joint-quantumcorrelation-error}
\Delta\Pi^{ab}(\alpha,\beta)&\equiv&  
P_{ee}(\alpha,\beta)-P_{eg}(\alpha,\beta)-P_{ge}(\alpha,\beta)
+P_{gg}(\alpha,\beta) - \langle \hat{\Pi}^{ab}(\alpha,\beta)\rangle
\nonumber \\
&=& - 2
  \langle\psi_f^{ab}|\hat{\Pi}^{ab}(\alpha,\beta)
  \left[\Delta(\hat{n}_a(\alpha))+\Delta(\hat{n}_b(\beta))\right]
  |\psi^{ab}\rangle 
\end{eqnarray}
where 
\begin{equation}
\label{Delta}
\Delta(\hat{n}_{a,b}(\alpha)) = {1 \over 4} 
[2\hat{n}_{a,b}(\alpha)+1]^2 \delta^2 +
{1\over 2}[2\hat{n}_{a,b}(\alpha)+1] \delta \Delta\phi + 
{1 \over 4}(\Delta\phi)^2
\end{equation} 
and $\hat{n}_{a,b}(\alpha) = \hat{D}^\dagger(\alpha)
\hat{n}_{a,b}\hat{D}(\alpha)$ are the displaced
number operators for the field modes in the cavities $a$ and $b$. 
The mean error $\Delta
B(\alpha,\beta)$ of the Bell function measurement in
(\ref{inequality1}) is given by
\begin{equation}
  \label{eq:fbf}
  \Delta B(\alpha,\beta) = \Delta\Pi^{ab}(0,0) + \Delta\Pi^{ab}(\alpha,0) +
  \Delta\Pi^{ab}(0,\beta) - \Delta\Pi^{ab}(\alpha,\beta).
\end{equation}

Consider an explicit example of a quantum nonlocal field 
(\ref{Meystre}) for an illustration of the
experimental errors.   For simplicity, we take the  phase
factor zero, {\it i.e.}, $\varphi=0$.  
We know from an earlier work  \cite{Banaszek} that
Bell's inequality is maximally violated with $B \sim -2.19$ 
when $\alpha=-\beta$ and
$|\alpha|^2\sim 0.1$.  Substituting
$|\Psi_f\rangle$ of Eq.~(\ref{Meystre}) into $|\psi_f^{ab}\rangle$
of Eq.~(\ref{joint-quantumcorrelation-error}), $\Delta\Pi^{ab}(\alpha,
\beta)$ is 
\begin{eqnarray}
  \label{eq:dpmfa}
  \Delta\Pi^{ab}(\alpha,\beta) &\approx& - 2
  \langle\psi^{ab}_f|\hat{\Pi}^{ab}(\alpha,\beta) |\psi_f^{ab}\rangle
  \langle\psi_f^{ab}|
  \left(\Delta(\hat{n}_a(\alpha))+\Delta(\hat{n}_b(\beta))\right) 
  |\psi_f^{ab}\rangle \nonumber \\
  &=& -2 \langle \hat{\Pi}^{ab}(\alpha,\beta) \rangle
  \left\{c_1(\alpha,\beta)\delta^2 + c_2(\alpha,\beta) \delta
  \Delta\phi + (\Delta\phi)^2 \right\}
\end{eqnarray}
where the mean field approximation has been used \cite{mean}.
The expectation value $\langle \hat{\Pi}^{ab}(\alpha,\beta) \rangle =
(2|\alpha-\beta|^2-1)e^{-2(|\alpha|^2+|\beta|^2)}$, and the
parameters 
$c_1(\alpha,\beta)=2(|\alpha|^4+|\beta|^4) +
\frac{13}{2}(|\alpha|^2+|\beta|^2) + 5$, and
$c_2(\alpha,\beta)=2(|\alpha|^2+|\beta|^2) + 4$. The mean error
$\Delta B$ for $\alpha=-\beta\approx \sqrt{0.1}$, is given by
\begin{equation}
  \label{eq:epefbt}
  \Delta B \sim 10.2\delta^2 + 8.1\delta\Delta\phi + 2.0(\Delta\phi)^2.
\end{equation}
The probing atoms normally have the Gaussian velocity distribution 
which causes the errors $\delta$ and $\Delta\phi$. When we consider
the ensemble average over atoms, the second term vanishes in 
Eq.~(\ref{eq:epefbt}) and the first and
third terms finally contribute to degrade the value of the Bell function.

For the test of nonlocality using single atoms as shown in Fig. 1(b), 
the mean error $\Delta B$ is similarly obtained as
\begin{equation}
  \label{eq:epefbo}
  \Delta B \sim 16.3\delta^2 + 8.2\delta\Delta\phi + 1.0(\Delta\phi)^2.
\end{equation}
This is slightly larger than the error (\ref{eq:epefbt}) 
for the two-atom scheme. The error
enhancement in the single-atom scheme is due to the  
fact that the experimental errors are multiplied as the
atom passes through two cavities.  For the two-atom scheme,
the error is a sum of errors occurred in each atom interaction with
the cavity field and $\pi/2$ pulses.

We find that when the standard deviation of the atomic velocity 
distribution is 5\%,
$\Delta B \sim 0.06$ for the two-atom scheme and $\Delta B
\sim 0.10$ for the single-atom scheme, which still allows
the observation of the violation of Bell's inequality.  

If the $Q$ factor of a cavity is high, the cavity is very much closed.
When an atom passes through the cavity walls the atom can lose the
information on the phases of atomic states so that the scheme suggested in
this paper cannot be used.  However, recently, Nogues {\it et al.}
suggested a way to implement $\pi/2$ pulses and cavity fields
interactions inside the cavity \cite{Nogues99}.  If this scheme is
applied there will not be a problem of losing the $Q$ value to keep
the atomic phase information.

\acknowledgements

M.S.K. thanks Professor Walther for discussions and hospitality at the
Max-Planck-Instut f\"{u}r Quantenoptik where a part of this work was
carried out.  This work was supported by the BK21 grant (D-0055) 
by the Korean
Ministry of Education.

\begin{figure}
  \begin{center}
    \caption{Schematic diagram of the quantum nonlocality test for 
      cavity fields. (a) Two two-level atoms pass through two cavities
      and joint measurements of atomic levels are performed after the
      cavity interaction.  (b) A single two-level atom passes 
      sequentially through two
      cavities and a measurement of atomic level is performed after
      the atom-field interaction in the cavities.}
  \end{center}
  \label{fig:configuration}
\end{figure}


\begin{references}
\bibitem{mean} We checked the mean field approximation as we calculate
  the exact value of Eq.~(\ref{joint-quantumcorrelation-error}) 
  for the nonlocal state
  (\ref{Meystre}).  The result in Eq.~(\ref{eq:epefbt}) is  
  within $0.1\%$ of the exact algebraic calculation which implies that 
  the correlation effect between $\hat{\Pi}$ and
  $\Delta(\hat{n})$ is negligible for fields of small amplitudes. 
  We conjecture that the mean field approximation holds for the 
  displaced nonlocal fields of the displacements 
  $|\alpha| < 1$ and $|\beta| < 1$.
\end{references}
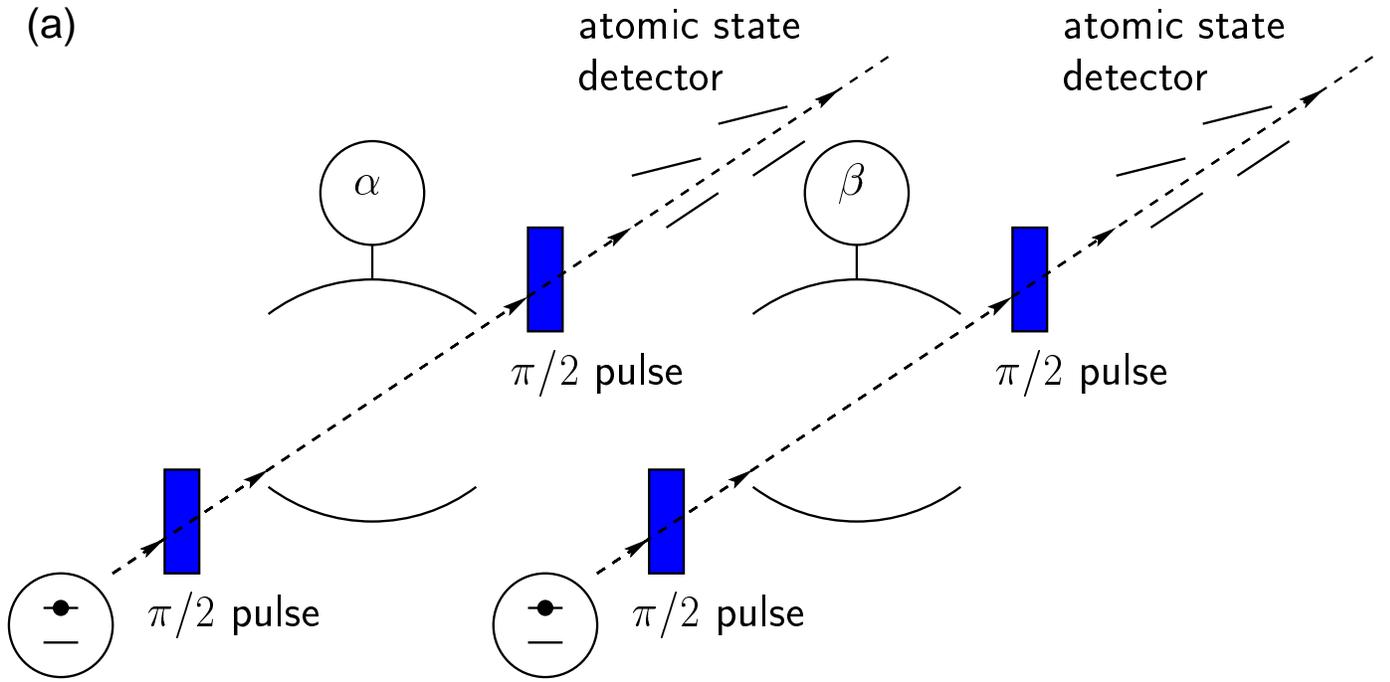
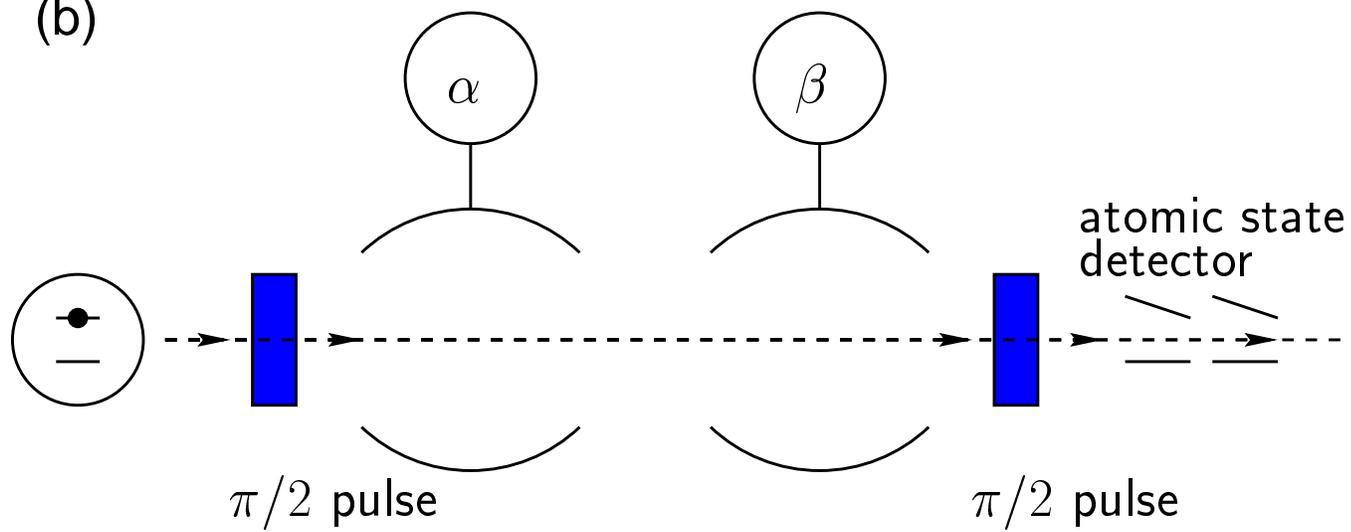
\end{document}